\documentclass[conference]{IEEEtran}
\IEEEoverridecommandlockouts
\usepackage{cite}
\usepackage{amsmath,amssymb,amsfonts}
\usepackage{algorithmic}
\usepackage{graphicx}
\usepackage{enumitem}
\usepackage{textcomp}
\usepackage{xcolor}
\usepackage[scientific-notation=true]{siunitx}
\usepackage{subcaption}
\usepackage{etoolbox} 

\usepackage{amsmath,amsfonts,bm,amssymb}

\def\eqref#1{equation~\ref{#1}}

\def\1{\bm{1}}

\def\vw{{\bm{w}}}
\def\vx{{\bm{x}}}
\def\vy{{\bm{y}}}

\DeclareMathAlphabet{\mathsfit}{\encodingdefault}{\sfdefault}{m}{sl}
\SetMathAlphabet{\mathsfit}{bold}{\encodingdefault}{\sfdefault}{bx}{n}

\usepackage{tikz}
\usepackage{textcomp}
\usepackage{hyperref}

\def\fnet{{\Phi}}

\def\fnetp{{\fnet(\vx;\theta)}}

\def\BibTeX{{\rm B\kern-.05em{\sc i\kern-.025em b}\kern-.08em
    T\kern-.1667em\lower.7ex\hbox{E}\kern-.125emX}}

\newcommand\copyrighttext{%
\footnotesize \textcopyright 2021 IEEE. Personal use of this material is permitted.
Permission from IEEE must be obtained for all other uses, in any current or future
media, including reprinting/republishing this material for advertising or promotional
purposes, creating new collective works, for resale or redistribution to servers or
lists, or reuse of any copyrighted component of this work in other works.}
\newcommand\copyrightnotice{%
\begin{tikzpicture}[remember picture,overlay]
\node[anchor=south,yshift=10pt] at (current page.south) {\fbox{\parbox{\dimexpr\textwidth-\fboxsep-\fboxrule\relax}{\copyrighttext}}};
\end{tikzpicture}%
}

\begin{document}

\title{Learning Free-Surface Flow with Physics-Informed Neural Networks\\
}



\author{\IEEEauthorblockN{Raphael Leiteritz, Marcel Hurler, Dirk Pfl\"{u}ger}
\IEEEauthorblockA{Institute of Parallel and Distributed Systems\\
Department of Scientific Computing\\
University of Stuttgart\\
Stuttgart, Germany\\
Email: \{raphael.leiteritz, hurlerml, dirk.pflueger\}@ipvs.uni-stuttgart.de}
}

\maketitle
\copyrightnotice

\begin{abstract}
    The interface between data-driven learning methods and classical simulation poses an interesting field offering a multitude of new applications.
    In this work, we build on the notion of physics-informed neural networks (PINNs) and employ them in the area of shallow-water equation (SWE) models.
    These models play an important role in modeling and simulating free-surface flow scenarios such as in flood-wave propagation or tsunami waves.
    Different formulations of the PINN residual are compared to each other and multiple optimizations are being evaluated to speed up the convergence rate.
    We test these with different 1-D and 2-D experiments and finally demonstrate that regarding a SWE scenario with varying bathymetry, the method is able to produce competitive results in comparison to the direct numerical simulation with a total relative $L_2$ error of $\num{8.9e-3}$.
    
\end{abstract}


\section{Introduction}

When modeling fluid flow problems, typical methods of choice are Finite Elements, Finite Volumes or similar numerical discretization schemes.
However, with the advent of scientific machine learning (SciML) a new approach of solving such problems has gained a lot of traction in recent years.
In the wake of the machine learning revolution, more and more research is being conducted at what can be described as the interface of data-driven methods and classical approaches.
 While in some works only the input-output mapping from experimental data is considered and fed into a black-box neural network (NN) \cite{nn_pde_data}, others focus on using special architectures such as long-short term memory (LSTM) NNs to capture temporal effects \cite{dl_lstm} or convolutional neural networks (CNNs) \cite{nn_pde_cnn} in order to extract spatial features.
 Another class of neural network based approaches is to impose soft constraints during the training process of NNs to inject additional information.
 This idea has already been studied quite a while back e.g. in \cite{nn_pde1, nn_pde2} by using the underlying partial differential equations (PDEs) as source of physical information.
 This, however, was heavily inhibited by computational limitations because gradients in the loss function had to be computed carefully by hand.
 Due to the recent advancements in machine learning, this original idea has now been revived under the notion of physics-informed neural networks by \cite{pinn} and is being adopted quickly.
 Modern software frameworks such as PyTorch \cite{pytorch} or TensorFlow \cite{tensorflow} function as readily available tools to implement and solve such approaches by offering powerful automatic differentiation capabilities.
 Since then, PINNs have been extended to uncertain regimes \cite{pinn_uq} and applied to various different problem setups such as fluid mechanics \cite{pinn_hfm} or subsurface flows \cite{pinn_subsurface}.

 In this paper, we employ PINNs to solve free-surface flow problems which are described by the shallow water equations. We develop a new scalar-valued loss function which fulfills mass conservation by design and compare this with two alternatives. Furthermore, we evaluate multiple training optimizations for PINNs which have been proposed in recent publications.

 \section{Prerequisites}
 \subsection{Physics-Informed Neural Networks}
 To build a physics-informed neural network we follow the notation as described in \cite{pinn}.
 Starting from a single layer
 \begin{equation}
   f(\vx) = \sigma(\vx^\intercal \vw + b) \, ,
 \end{equation}
 where $\vx \in \mathbb{R}^d$ are the network inputs, $\vw \in \mathbb{R}^d$ are learnable weights, $b$ the bias and $\sigma : \mathbb{R} \rightarrow \mathbb{R}$ some non-linear activation function.
 Compositing a number $n$ of these layers gives a simple fully-connected feed-forward network
 \begin{equation}
   \Phi(\vx, \theta) = (f_n \circ \cdots \circ f_2 \circ f_1)(\vx)
 \end{equation}
 with $\theta$ consisting of all trainable weight parameters.
 We optimize this neural network w.r.t its parameters by minimizing the loss function
 \begin{equation}
   \label{eq:loss}
   \mathcal{L}(\theta) = \frac{1}{N} \sum_{i=1}^N \left( \vy_i - \fnet(\vx_i; \theta) \right)^2
 \end{equation}
 in order to minimize the mean squared error of the network with respect to some true data points $\{\vx_i, \vy_i\}_{i=1}^N$.

 Furthermore, we assume that we have a problem at hand whose underlying dynamics can be described by an arbitrary PDE of the form
 \begin{align}
    \mathcal{N}_{\vx}\left[u\right] &= 0, \qquad \quad\,\vx \in \Omega \subset \mathbb{R}^d \\
    u &= z(\vx_z), \quad \vx_z \in \partial \Omega, z : \partial \Omega \rightarrow \mathbb{R}^n\,,
  \end{align}
 where $u : \Omega \rightarrow \mathbb{R}^n$ is the solution operator, $z$ is some function describing boundary conditions, and $\mathcal{N}_{\vx}$ is a generally non-linear differential operator.
 We use this ground truth to create a physics-informed neural network by adding a new physical loss term
 \begin{equation}
    \label{eq:phys_loss}
    \ell_{p}(\vx;\theta) = \mathcal{N}_{\vx}\left[\fnetp\right]^2 \, ,
 \end{equation}
 to the loss function.
 This penalizes the PDE residual with the current network prediction substituted for the true solution. 
 The resulting total physics-informed loss is defined as
 \begin{equation}
    \mathcal{L}_{p}(\theta) = \frac{1}{N} \sum_{i=1}^N \left( \vy_i - \fnet(\vx_i; \theta) \right)^2 + \ell_{p}(\vx;\theta) \, ,
 \end{equation}
 where the first part corresponds to a general mean-squared error (MSE) loss evaluated at some true data points $\{\vx_i, \vy_i\}_{i=1}^N$.

 \subsection{Shallow Water Equations}
 \label{sec:swe}
\begin{figure}[hbt]
\centering
    \includegraphics[width=0.8\linewidth]{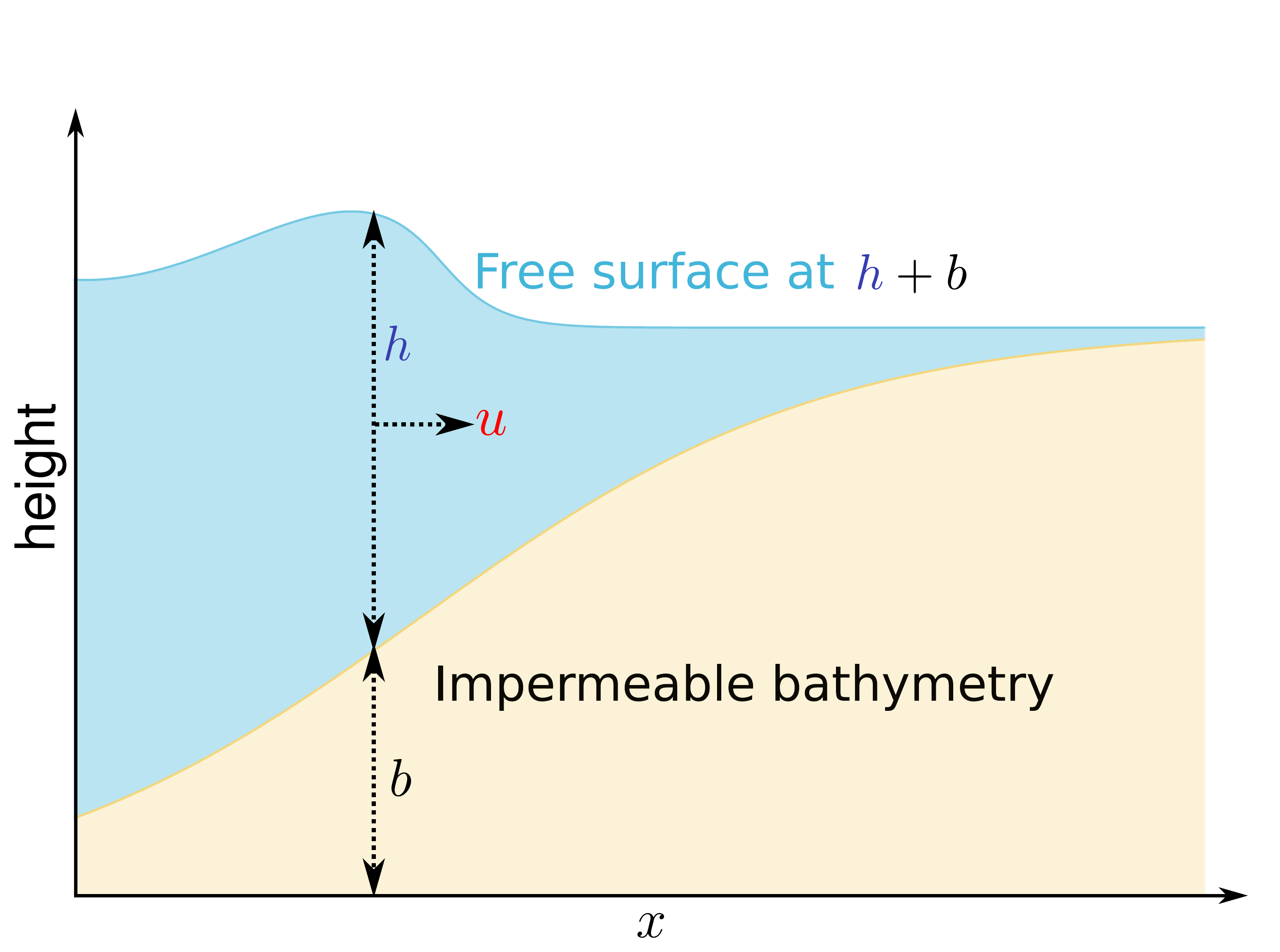}
    \caption{Illustration of the simulation setup including relevant variables.}
    \label{fig:setup}
\end{figure}

The shallow-water equations, a simplification of the general Navier-Stokes equations, present a suitable approximation to model free-surface flow problems \cite{swe_navier}.
  They are used in many applications, such as simulating and predicting tsunami waves or large flooding events \cite{tsunami}.
  In this work, we focus on the shallow water equations which come in the form of the following system of hyperbolic PDEs
  \begin{align}
    \label{eq:swe1}
     \frac{\partial h}{\partial t} + \frac{\partial hu}{\partial x} + \frac{\partial hv}{\partial y} &= 0 \\
     \label{eq:swe2}
     \frac{\partial hu}{\partial t} + \frac{\partial u^2h + \frac{1}{2} g_r h^2}{\partial x} &= - g_r h \frac{\partial b}{\partial x} \, , \\
     \label{eq:swe3}
     \frac{\partial hv}{\partial t} + \frac{\partial v^2h + \frac{1}{2} g_r h^2}{\partial y} &= - g_r h \frac{\partial b}{\partial y} \, ,
  \end{align}
  where $u$ and $v$ are the velocities in horizontal and vertical direction respectively, $h$ is the water depth and $hu$ as well as $hv$ can be interpreted as the directional momentum components.
  Additionally, $g_r$ represents a gravitational force and $b(x)$ a function describing the bathymetry.
  The first equation is derived from conservation of mass while the latter two are derived from the conservation of momentum.
 Friction terms are not considered in this setting.
 A 1-D illustration of the experimental setup is shown in Figure \ref{fig:setup}.

  In order to set up a baseline and to generate validation data, the scenarios described in Section \ref{sec:results} are first simulated numerically.
  A Finite-Volume discretization is employed using a Godunov flux method for solving the resulting Riemann problems at the control volume interfaces.
 This numerical reference simulation was implemented by using the Clawpack package \cite{clawpack} as it offers a collection of finite volume methods for linear and non-linear hyperbolic systems as well as an easy-to-use Python interface called pyClaw \cite{pyclaw-sisc}.
 As suggested by the Clawpack documentation, for cases with a flat bathymetry the Riemann problem is approximated using the HLLE method \cite{hlle} while for scenarios with varying bathymetry an $f$-wave approach \cite{fwave} is chosen.

\section{Model}
\subsection{A PINN SWE Model in 1D}
\label{sec:model}
Given the equations \ref{eq:swe1} \& \ref{eq:swe2}, the straightforward approach for modelling SWEs in a PINN context would be to let the network $\Phi(x,t;\theta)$ learn the water height and momentum $\tilde{h}(x,t), \widetilde{hu}(x,t)$ directly, resulting in the combined physical loss
\begin{equation}
   \ell_p^{C} = \left[ \frac{\partial \tilde{h}}{\partial t} + \frac{\partial \widetilde{hu}}{\partial x} \right]^2 + \left[\frac{\partial \widetilde{hu}}{\partial t} + \frac{\partial \widetilde{hu}\tilde{u} + \frac{1}{2} g_r \tilde{h}^2}{\partial x} + g_r \tilde{h} \frac{\partial b}{\partial x} \right]^2 \, .
\end{equation}
However, since we do not have a direct representation for the velocity $u$ we have to approximate it by $\tilde{u} = \frac{\widetilde{hu}}{\tilde{h}}$.
This poses the danger of numerical errors stemming from zero- or close-to-zero-divisions since we can not guarantee the network to make non-zero predictions for $\tilde{h}$ especially in the beginning of the training.
Alternatively one can also rewrite the SWEs in the non-conservative form and build a PINN that directly outputs height and velocity $\tilde{h}(x,t), \tilde{u}(x,t)$
\begin{equation}
   \ell_p^{NC} = \left[ \frac{\partial \tilde{h}}{\partial t} + \frac{\partial \tilde{h}\tilde{u}}{\partial x} \right]^2 + \left[\frac{\partial \tilde{h}\tilde{u}}{\partial t} + \frac{\partial \tilde{h}\tilde{u}^2 + \frac{1}{2} g_r \tilde{h}^2}{\partial x} + g_r \tilde{h} \frac{\partial b}{\partial x} \right]^2 \,
\end{equation}
which eliminates this issue altogether.
Thirdly, the equations may as well be rewritten in a form that allows for the mass equation \ref{eq:swe1} to be fulfilled by definition.
First, a scalar-valued function $\phi : \Omega \rightarrow \mathbb{R}$ is considered that is at least twice continuously differentiable.
We know that by definition $\text{curl}\, \nabla \phi$ must be zero.
If we now define $\tilde{h} := -\frac{\partial \phi}{\partial x}$ and $\widetilde{hu} := \frac{\partial \phi}{\partial t}$, we can simplify the PINN loss to
\begin{equation}
   \ell_p^{\phi} = \left[\frac{\partial^2 \tilde{\phi}}{\partial t^2} + \frac{\partial \tilde{\phi}_t^2 + \frac{1}{2} g_r -\tilde{\phi}_x^2}{\partial x} + g_r \tilde{h} \frac{\partial b}{\partial x} \right]^2 \, ,
\end{equation}
given that the PINN now outputs a prediction of the scalar valued function $\tilde{\phi}$.

We compare all three approaches for a simple depth perturbation scenario characterized by a flat bathymetry and the initial water height and velocity given as
\begin{align*}
   &h(x,0) = 0.2 \exp(\frac{-x^2}{0.4}) + 0.8 \\
   &u(x,0) = 0 \, .
\end{align*}

\begin{figure}[hbt!]
    \centering
    \includegraphics[width=\linewidth]{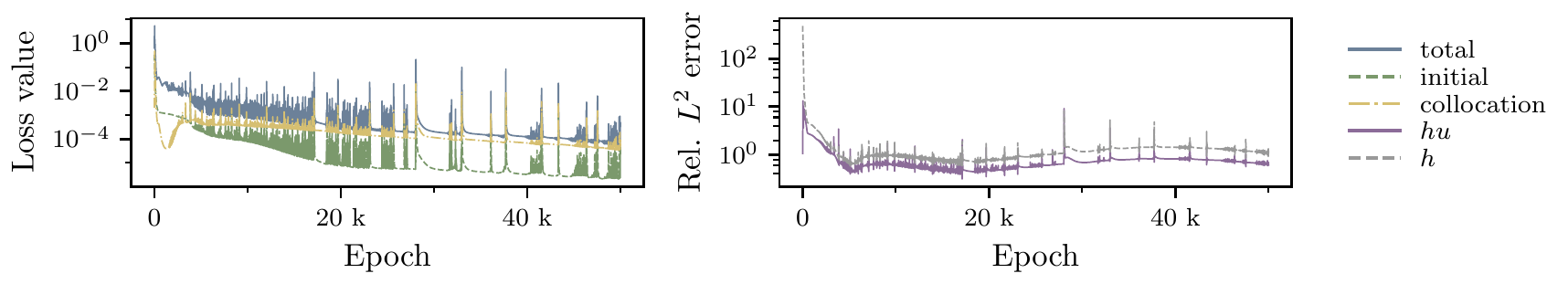}
    \caption{Training losses and predictions of $h$ and $u$ plotted over 50k epochs for the simple depth perturbation scenario using the scalar function approach.}
    \label{fig:scalar}
\end{figure}

\begin{figure}[hbt]
    \centering
    \includegraphics[width=\linewidth]{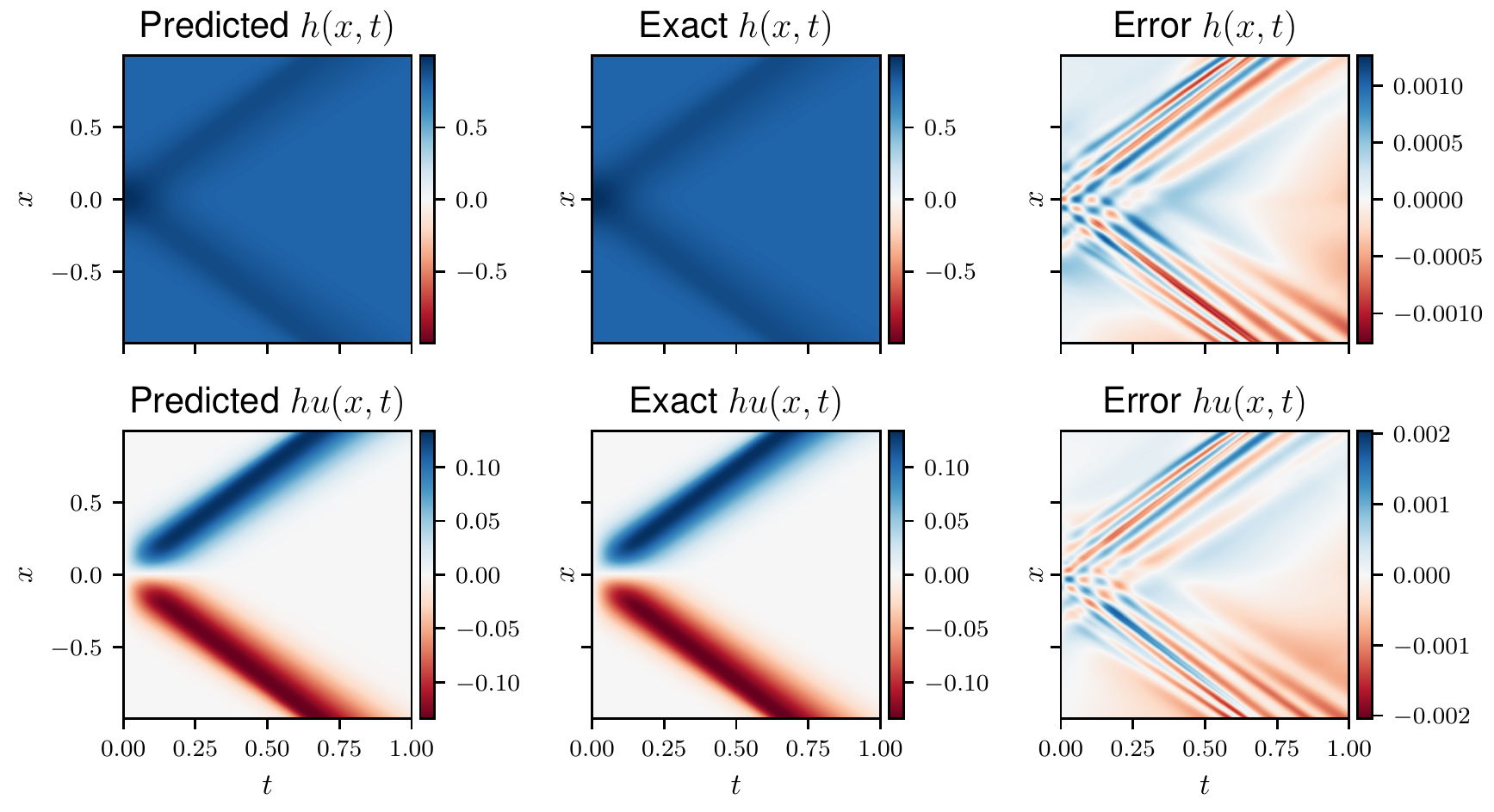}
    \caption{Comparison of predicted and simulated solution for the simple depth perturbation scenario using the non-conservative model.}
    \label{fig:simple_prediction}
\end{figure}

The results of this initial benchmark for all three approaches are presented in Table \ref{tbl:pinn_choice}.
As can be seen clearly, both the non-conservative and conservative form achieve a reasonably good prediction error after $50$K training epochs.
Surprisingly however, even though the scalar-valued approach satisfies mass conservation by definition, this does not automatically equal a better approximation.
The validation graph in Figure \ref{fig:scalar} clearly shows that the performance stagnates after just $8$K epochs, indicating some sort of overfitting happening.
Utilizing these results, we decided to only run the non-conservative PINN as our architecture of choice for the following scenarios.
The corresponding approximation results for the best method are exemplary shown in Figure \ref{fig:simple_prediction}.


\sisetup{detect-weight,mode=text}
\renewrobustcmd{\bfseries}{\fontseries{b}\selectfont}
\renewrobustcmd{\boldmath}{}
\newrobustcmd{\B}{\bfseries}

\begin{table}[th]
   \centering
   \caption{Depth Perturbation Scenario --- Performances of different PINN types, each row corresponds to the network with the lowest total validation error achieved after training 50K epochs.}
   \begin{tabular}{l|ccccc}
     PINN Type & total err.     & $h$ err.       & $hu$ err.      & mass eq.       & mom.\ eq.\\
     \hline
     $\ell_p^{\mathrm{C}}$               & \num{2.8e-04} & \num{3.2e-04} & \num{2.4e-04} & \num{2.5e-06} & \num{2.7e-06}\\
     $\ell_{p}^\phi$                     & \num{5.5e-01} & \num{6.7e-01} & \num{4.4e-01} & \num{0.0}        & \num{4.1e-04}\\
     $\ell_{p}^{\mathrm{NC}}$   & \num{1.2e-04} & \num{1.3e-04} & \num{1.1e-04} & \num{3.1e-06} & \num{2.3e-06}
   \end{tabular}
   \label{tbl:pinn_choice}
 \end{table}

\subsection{Training Optimizations}

In order to speed up the training and to improve the prediction accuracy, we evaluate a few different strategies that have been proposed in various PINN-related works.
To weight the contribution of the different parts of the physical loss we utilize the learning-rate annealing (LRA) method as proposed by \cite{lra}.
This will be used throughout all of our test runs to establish a baseline.

Additionally, we explore the use of three more optimization methods:

\begin{description}[style=unboxed,leftmargin=0cm]
  \item[Sinusoidal Activation Functions] By using sine activation functions in their networks, the work of \cite{siren} has shown some nice accuracy gains of over an order of magnitude for a wave equation PINN model.
  This is not only attributed to the frequency representation that sinusoidal activation functions allow for, but also to a specialized initialization scheme which ensures that the training process is stable.
  However, this method also introduces a new hyper-parameter $\kappa$ that controls the activation function's frequency with respect to the input domain of each layer.
  This had to be hand-tuned to a value of $\kappa=3.5$ to work well with our test cases as opposed to the author's suggestion of keeping it fixed at $\kappa=30$.

  \item[Locally Adaptive Activation Functions] Similarly, \cite{laaf} introduced a scaling factor on a per-layer (L-LAAF) or per-neuron (N-LAAF) basis for the network's activation functions.
  These factors, however, are treated as additional trainable parameters of the network and are thus optimized during the normal training process.
   
   \item[Attention Mechanism]Another promising way of increasing the network's accuracy was introduced by \cite{McClenny2020}.
   In their work, they postulate that stiff PDEs, which manifest themselves in sharp interfaces in the spatio-temporal domain, are typically hard to train using the standard PINN approach and therefore require a lot of collocation points to resolve the mentioned interfaces.
   They try to mitigate this by introducing an attention mechanism for the collocation points, which all get weighted individually in the physical loss.
   The optimization problem is  reformulated into a saddle point problem \(\min_{\theta} \max_{\mathbf{\lambda}} \mathcal{L}_{p}(\theta, \mathbf{\lambda})\) which first maximizes the loss w.r.t.\ individual collocation point weights $\lambda_i$ and then minimizes it w.r.t.\ the usual network parameters $\theta$.
   This drives the optimizer's ``attention'' towards those areas that experience a high physical loss.
\end{description}

\section{Results}
\label{sec:results}
\subsection{Varying Bathymetry --- Sandbank Scenario}
\begin{figure}[hbt!]
   \centering
      \includegraphics[width=1.0\linewidth]{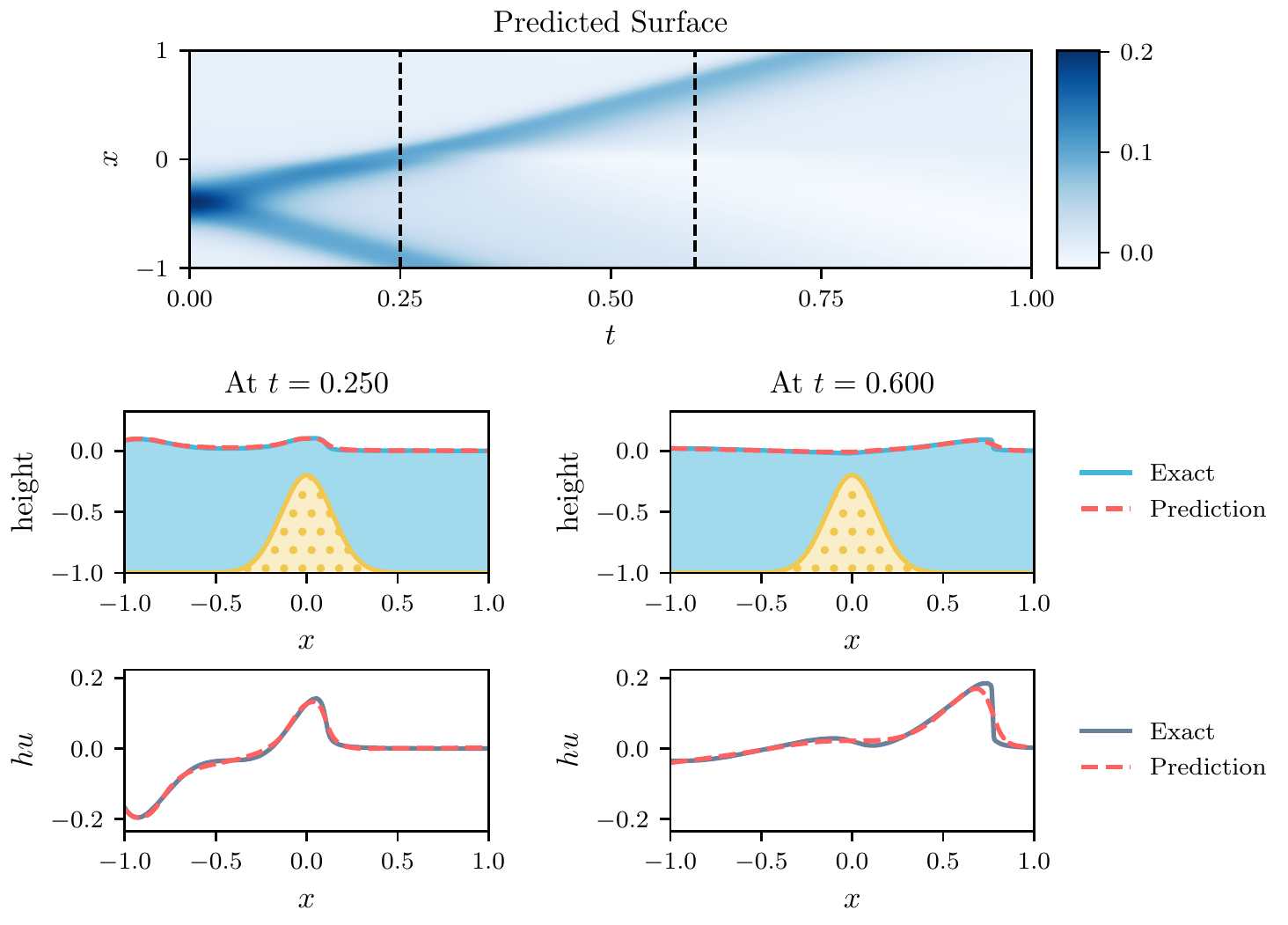}
      \caption{Visualization of the best model prediction for the sandbank scenario. Top: Overall solution in the time-space domain. Bottom: Detail plots of both height $h$ and momentum $hu$ prediction at two different time-steps.}
      \label{fig:sandbank}
 \end{figure}

The given methods are now applied to a more complex one-dimensional example where an initial depth perturbation of the fluid surface as well as a varying bathymetry are given.
The complete initial and boundary conditions are defined as
\begin{align*}
   &b(x) \quad = 0.8 \exp(\frac{-x^2}{0.4}) - 1 \\
   &h(x,0) = 0.2 \exp(\frac{-(x+0.4)^2}{0.4}) - b(x) ,\qquad u(x,0) = 0 \, ,
\end{align*}
and the setup is illustrated in Figure \ref{fig:sandbank}.

The hyper-parameters used for training the PINN in this scenario are chosen as follows:
The network is $5$ layers deep and $50$ neurons wide using \textit{tanh} activation functions.
This choice was found by hyperparameter optimization using a standard MSE loss against full simulation results obtained from the reference solution.
We apply standard input and weight normalization methods as suggested in the supplementary material of \cite{pinn_hfm}.
The LRA parameter is set as $\alpha_{lra} = 0.9$. For optimization a standard ADAM algorithm \cite{adam} is used with a learning rate of $\num{0.001}$.

\begin{figure}[hbt!]
   \centering
      \includegraphics[width=\linewidth]{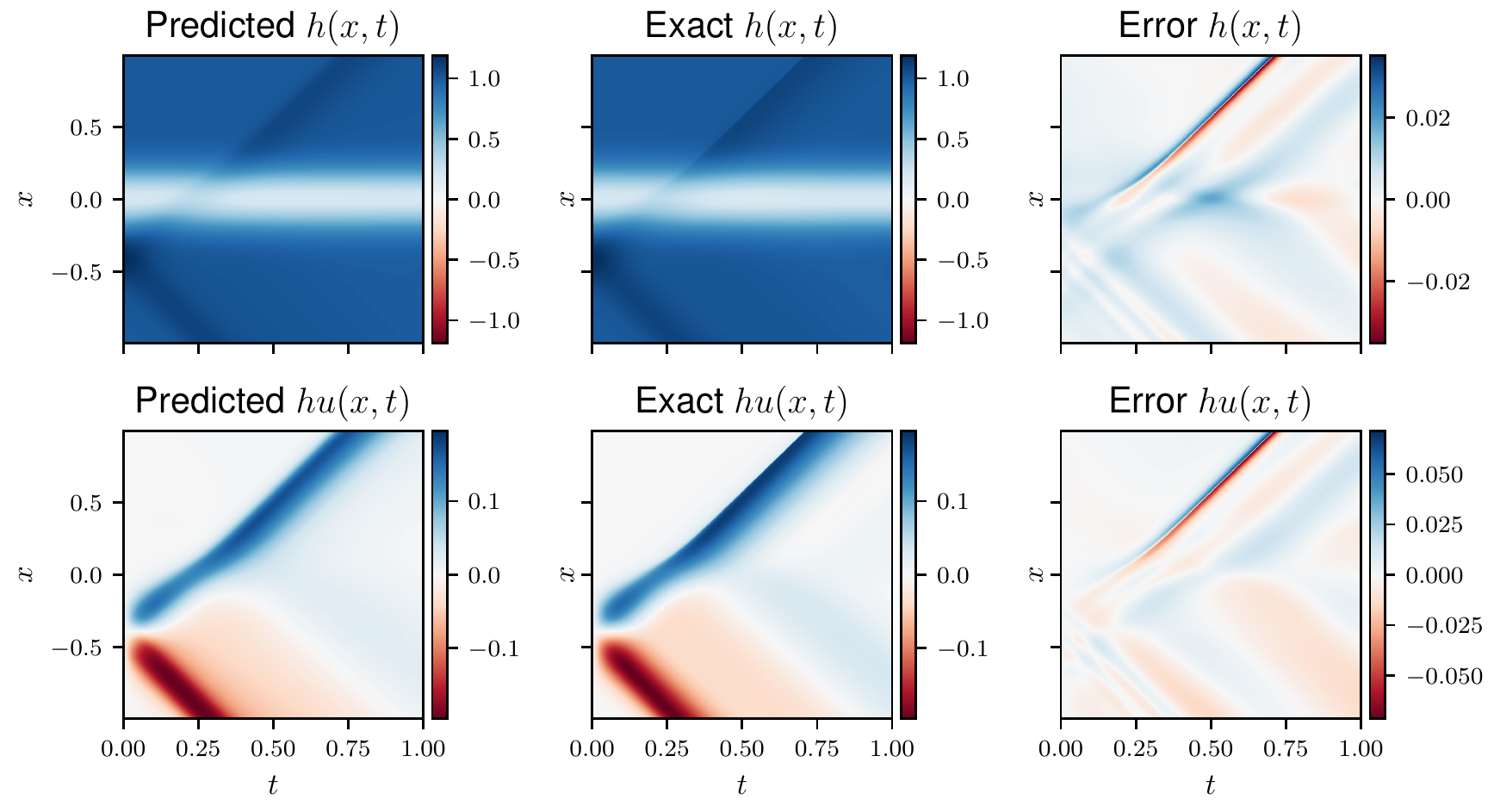}
      \caption{Comparison of predicted and simulated solution for the sandbank scenario using the N-LAAF method.}
      \label{fig:bathy_prediction}
\end{figure}

Figure \ref{fig:bathy_prediction} shows a comparison of the best PINN prediction versus the Finite Volume baseline.
This result was achieved using the N-LAAF method with a total relative L2-Error of just $\num{8.9e-03}$ after training for $50$k epochs.
We can clearly observe that the characteristics of the true solution are accurately predicted.
Both waves traveling in opposite directions of the domain are accurately resolved as well as the steepening of the wave traveling to the right of the domain while crossing over the sandbank bathymetry.
Additionally, we see from the time-space diagram that the error in both $h$ and $u$ is mostly concentrated on the shock that is forming at the wavefront.

\begin{table}[hbt!]
   \centering
   \caption{Overview of different optimization methods for the sandbank scenario after training 50K epochs. Shown are the total validation error as well as errors concerning the individual velocity, momentum predictions as well as the mass and momentum equation residuals.}
   \begin{tabular}{l|ccccc}
     Method               & total err.     & $h$ err.       & $hu$ err.      & mass eq.       & mom. eq.\\
     \hline
     LRA       & \num{2.4e-02} & \num{1.6e-03} & \num{4.6e-02} & \num{6.3e-04} & \num{3.5e-03}\\
     Siren                 & \num{1.1e-01} & \num{3.9e-03} & \num{2.1e-01} & \num{8.3e-04}        & \num{2.8e-03}\\
     L-LAAF      & \num{3.0e-02} & \num{1.6e-03} & \num{5.8e-02} & \B \num{1.0e-04} & \B \num{4.0e-04}\\
     N-LAAF      & \B \num{8.9e-03} & \B \num{4.1e-04} & \B \num{1.7e-02}  & \num{2.6e-04} & \num{4.5e-04}\\
     Attention             & \num{1.8e-02} & \num{1.2e-03} & \num{3.5e-02}  & \num{4.6e-03} & \num{4.4e-02}\\
   \end{tabular}
   \label{tab:sandbank}
 \end{table}

Table \ref{tab:sandbank} presents a more granular analysis of how the different optimization methods performed.
For this scenario, it is easy to spot that except for the neuron-wise adaptive activation function approach all others optimization methods failed to deliver a substantial improvement in prediction power.
Even after fine-tuning $\kappa$, the sine approach actually delivers an error which is an order of magnitude higher than the LRA baseline.

\begin{figure}[hbt]
    \centering
    \begin{subfigure}[b]{\linewidth}
       \caption{LRA}
       \includegraphics[width=\linewidth]{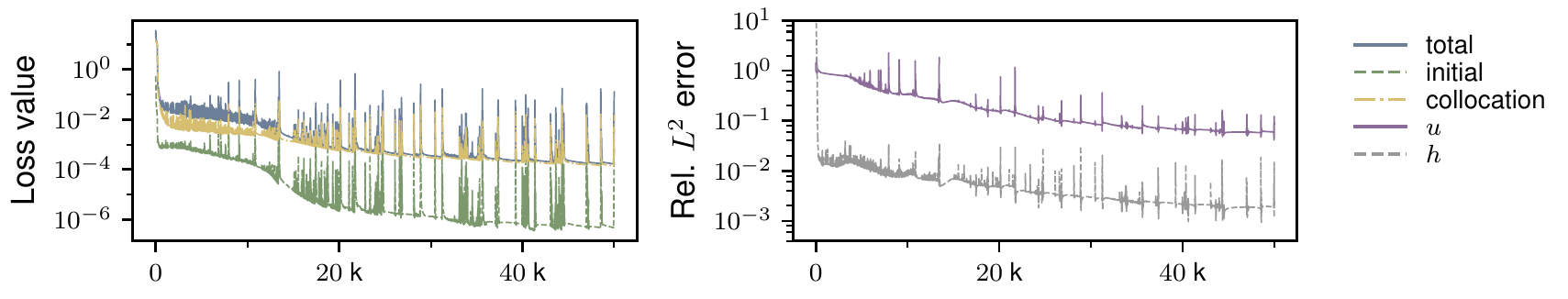}
       \label{fig:lra}
    \end{subfigure}
    \begin{subfigure}[b]{\linewidth}
       \caption{Attention}
       \includegraphics[width=\linewidth]{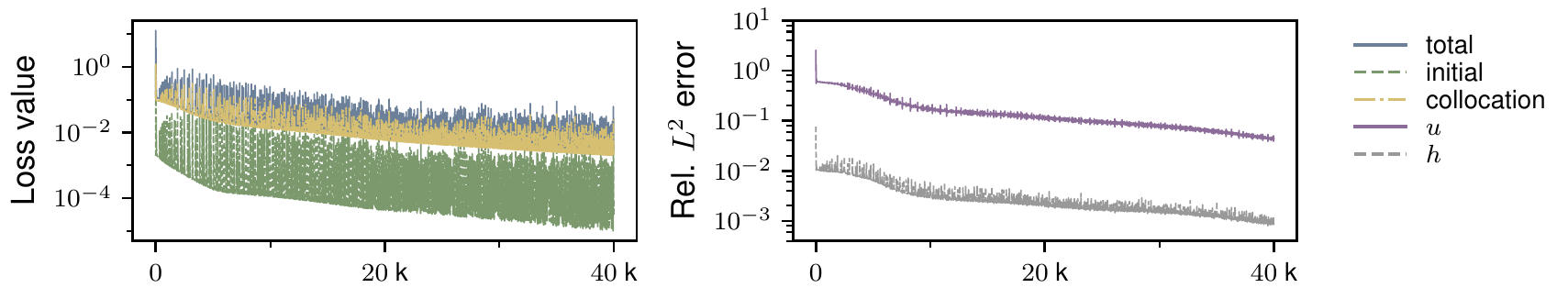}
       \label{fig:attention}
    \end{subfigure}
    \begin{subfigure}[b]{\linewidth}
       \caption{L-LAAF}
       \includegraphics[width=\linewidth]{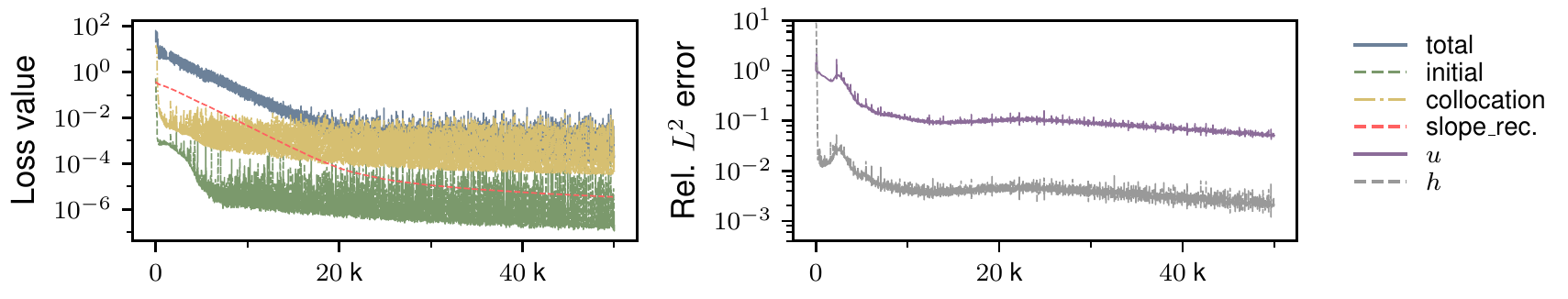}
       \label{fig:llaaf}
    \end{subfigure}
    \begin{subfigure}[b]{\linewidth}
       \caption{N-LAAF}
       \includegraphics[width=\linewidth]{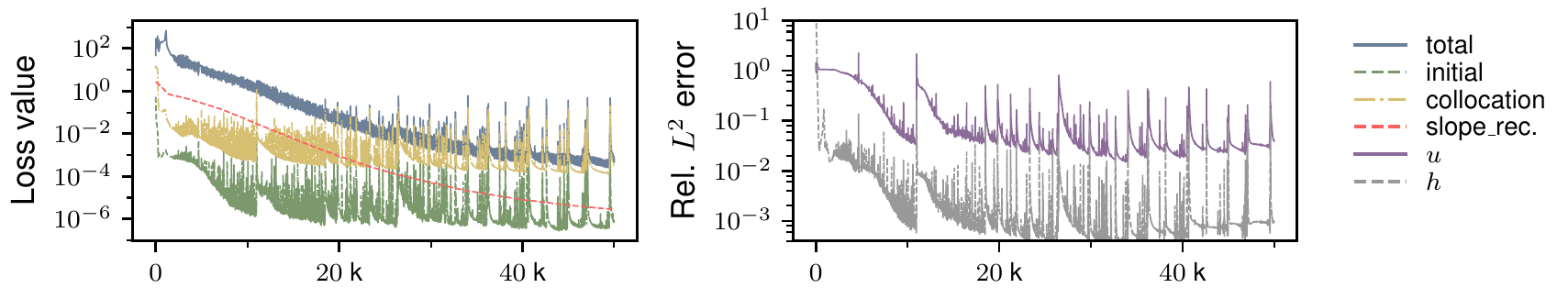}
       \label{fig:nlaaf}
    \end{subfigure}
    \begin{subfigure}[b]{\linewidth}
       \caption{Siren}
       \includegraphics[width=\linewidth]{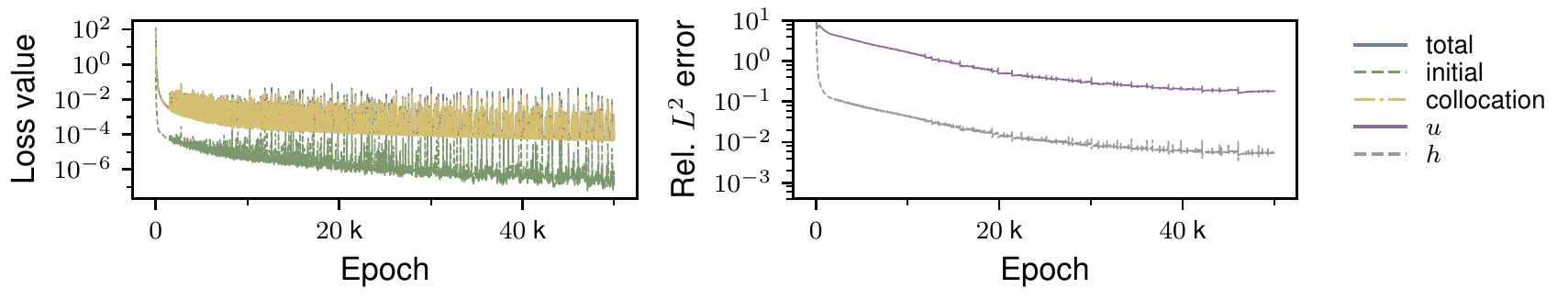}
       \label{fig:siren}
    \end{subfigure}
    \caption{Sandbank Scenario: Train losses and relative errors of $h$ and $u$ plotted over 50k epochs.}
    \label{fig:losses}
 \end{figure}

To examine this further we take a look at the evolution of training and validation metrics as plotted in Figure \ref{fig:losses}.
Here, we observe that while the adaptive activation function approaches do perform quite well in the overall relative errors, they seem to fall within an oscillatory regime which might be unfavorable.
In order to mitigate this behavior, it may be sufficient to do a hyper-parameter optimization with respect to the learning rate.
We can observe in general that the relative error of the momentum is usually one to two orders of magnitude higher than the relative height error.
This coincides with the previous observation, that most errors are close to steeper regimes of the height field.

\subsection{Radial Dam Break --- A 2D Scenario}
We now study a more challenging and realistic simulation scenario by extending the non-conservative approach as introduced in \ref{sec:model} to two dimensions.
Specifically, we consider a radial dam break scenario, where our initial condition is given by a circular bump of radius $r$ in an otherwise flat, square basin.
The boundary is modeled with outflow conditions allowing the wave to travel out of the domain.
The precise initial and boundary conditions are defined as
\begin{equation}
   h(x, y, 0) =
   \begin{cases}
      2,& \text{if } \sqrt{x^2 + y^2}  \geq r\\
      1,              & \text{otherwise}
   \end{cases}
\end{equation}
\begin{equation}
\begin{alignedat}{3}
   &\nabla h(x_b, y_b, t) \cdot \mathbf{n}  &&= \mathbf{0} \\
   &\nabla hu(x_b, y_b, t) \cdot \mathbf{n}  &&= \mathbf{0} \\
   &\nabla hv(x_b, y_b, t) \cdot \mathbf{n}  &&= \mathbf{0} \\
\end{alignedat}
\end{equation}
with $(x_b, y_b) \in \partial \Omega $, $\mathbf{n} \perp \partial \Omega$ and the domain $\Omega$ chosen as $[-2.5, 2.5] \otimes [-2.5, 2.5]$.
The baseline FV simulation was again conduction using pyClaw with $500$ timesteps up to $T_{max}=1.0$ with the same settings as described in \ref{sec:swe}.

\begin{figure}[hbt!]
   \centering
      \includegraphics[width=\linewidth]{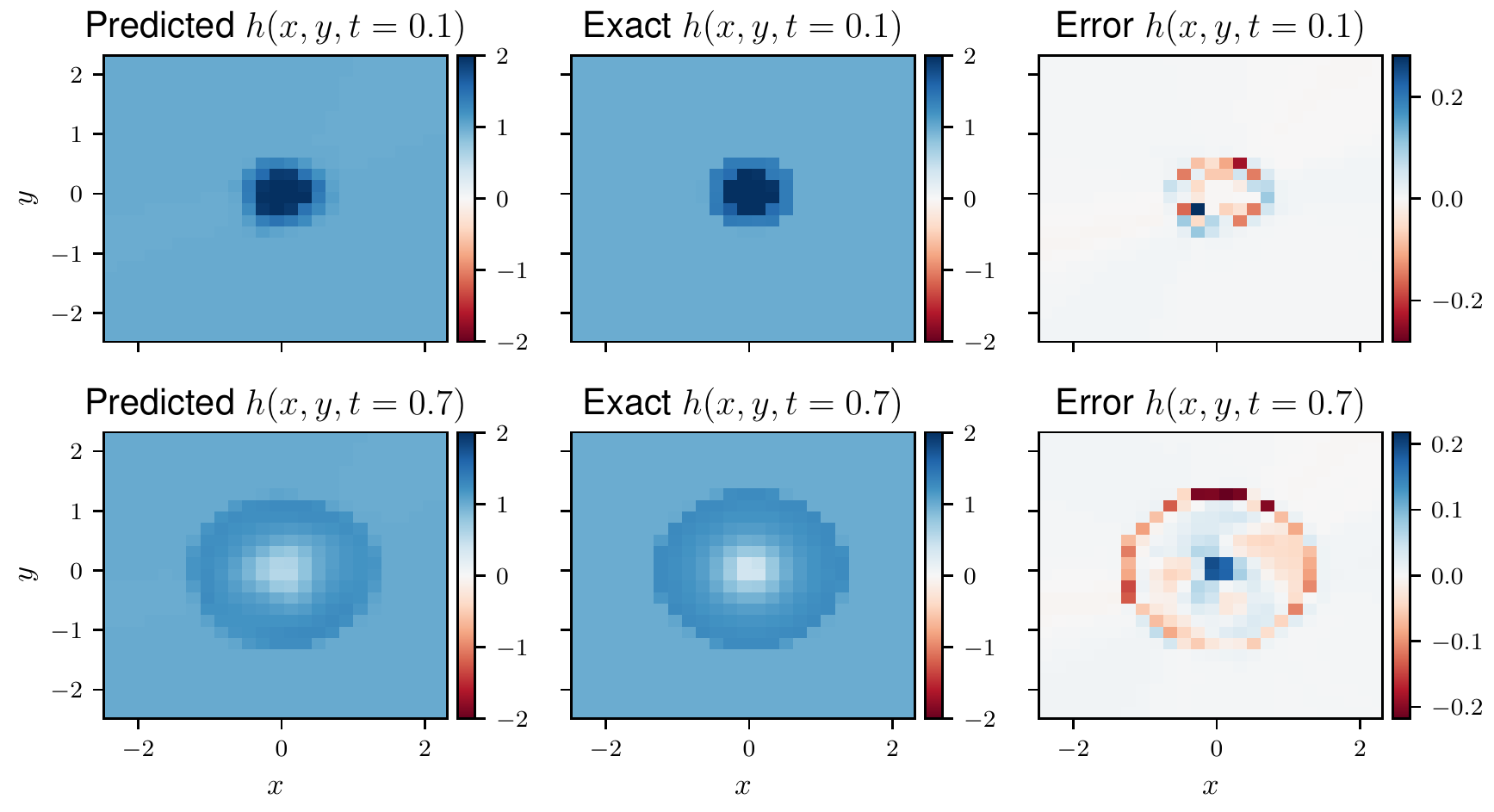}
      \caption{Comparison of predicted and simulated solution for the 2D radial dam break scenario using the LRA method.}
      \label{fig:2d_dam_prediction}
\end{figure}

Figure \ref{fig:2d_dam_prediction} qualitatively shows the network's learned water height prediction as well as the absolute error for two different timesteps $t=0.1$ and $t=0.7$.
As before, we can clearly observe that the characteristics of the simulation are captured by the network with the initial bump traveling outwards while slowly diffusing.
Most of the error in prediction is again focused around the steep wavefront which is traveling outwards from the center.

\begin{figure}[hbt]
    \centering
    \begin{subfigure}[b]{\linewidth}
       \caption{LRA}
       \includegraphics[width=\linewidth]{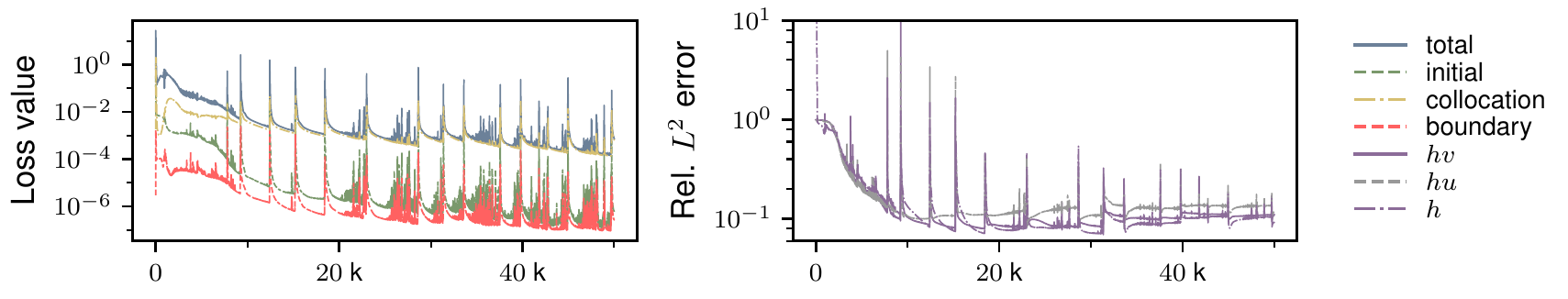}
       \label{fig:2d_lra}
    \end{subfigure}
    \begin{subfigure}[b]{\linewidth}
       \caption{Attention}
       \includegraphics[width=\linewidth]{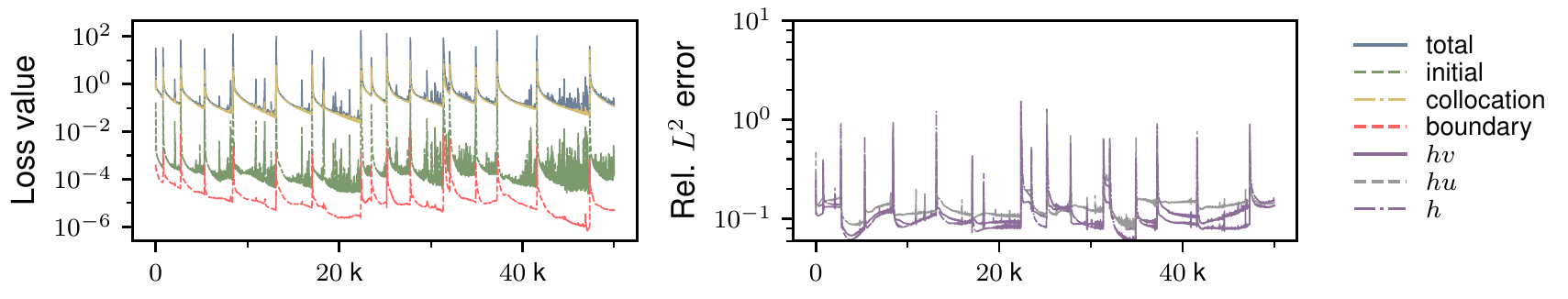}
       \label{fig:2d_attention}
    \end{subfigure}
    \begin{subfigure}[b]{\linewidth}
       \caption{L-LAAF}
       \includegraphics[width=\linewidth]{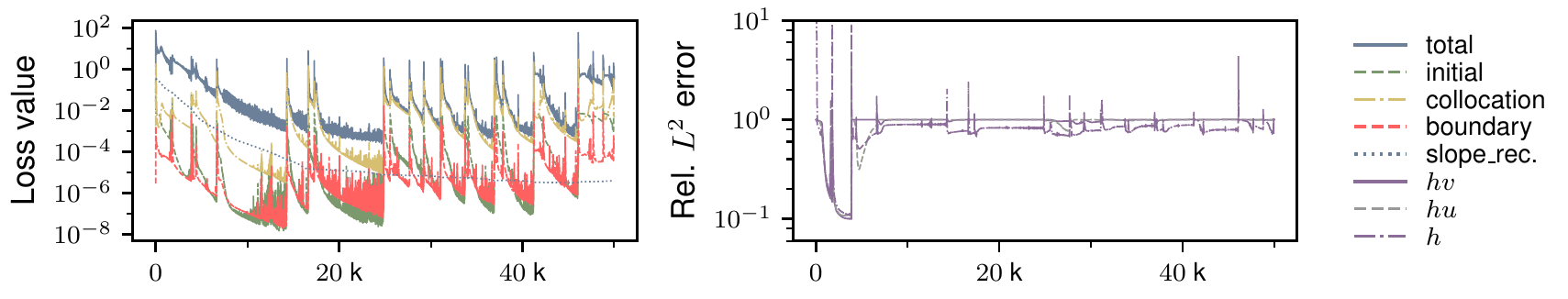}
       \label{fig:2d_llaaf}
    \end{subfigure}
    \begin{subfigure}[b]{\linewidth}
       \caption{N-LAAF}
       \includegraphics[width=\linewidth]{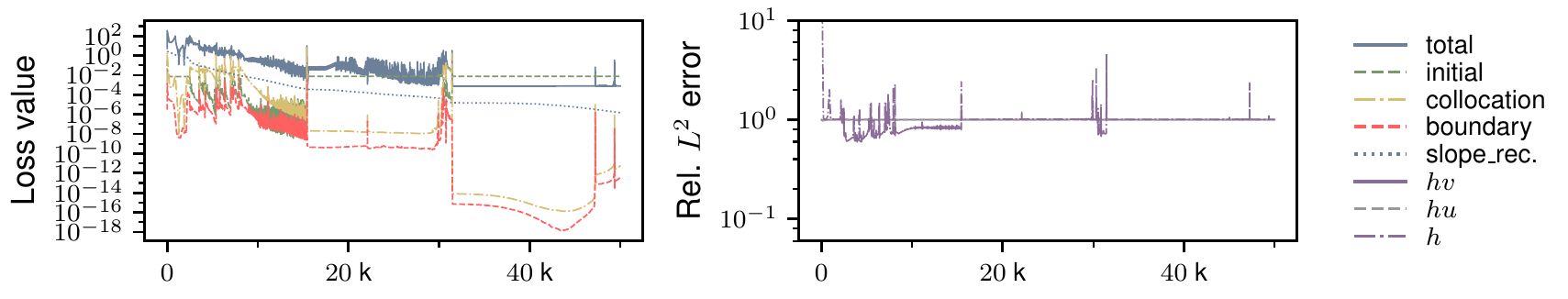}
       \label{fig:2d_nlaaf}
    \end{subfigure}
    \begin{subfigure}[b]{\linewidth}
       \caption{Siren}
       \includegraphics[width=\linewidth]{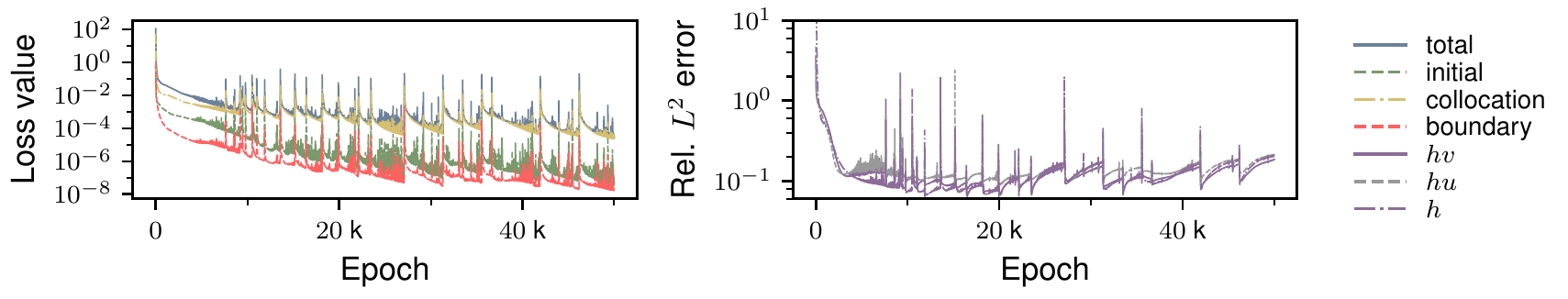}
       \label{fig:2d_siren}
    \end{subfigure}
    \caption{Radial Dam Break: Train losses and relative errors plotted over 50k epochs.}
    \label{fig:2d_losses}
 \end{figure}


 Again we present the best performance each optimization method achieved for this experiment in Table \ref{tab:2d_results}.
 For this scenario the LRA, Siren and Attention methods work more or less equally well with the best total relative error achieved at $\num{6e-2}$.
 The adaptive activation function approaches however, seem to fall behind.
 This is further demonstrated when looking at the individual loss plots in Figure \ref{fig:2d_losses}.
 It shows that both approaches not only show oscillatory behavior as in the application before but also become instable at some point during training and thus don't produce reasonable results anymore.
\begin{table}[hbt!]
    \centering
    \caption{Overview of different optimization methods for the 2D scenario after training 50K epochs. The total validation error is shown as well as errors concerning the individual velocity and momentum predictions.}
    \begin{tabular}{l|ccccc}
      Method               & total err.     & $h$ err.       & $hu$ err.      & $hv$ err. \\
      \hline
      LRA       & \num{8.3e-02}    & \num{7.5e-02}    & \num{1.0e-01}    & \num{6.9e-02}    \\
      Siren     & \num{7.4e-02}    & \num{7.8e-02}    & \num{7.9e-02}    & \B \num{6.5e-02}    \\
      L-LAAF    & \num{1.0e-01}    & \num{1.1e-01}    & \num{1.1e-01}    & \num{9.9e-02} \\
      N-LAAF    & \num{8.4e-01}    & \num{6.2e-01}    & \num{9.2e-01}    & \num{9.5e-01}    \\
      Attention & \B \num{7.2e-02} & \B  \num{6.3e-2} & \B\num{6.7e-02}  & \num{6.8e-02}    \\
    \end{tabular}
    \label{tab:2d_results}
\end{table}

According to these findings, we postulate that while the proposed optimization methods are performing well in their respective test cases, they also introduce more complexity in the form of new hyper-parameters which again need to be chosen carefully.
Otherwise, they might even pose the risk of performing worse than the baseline LRA method.

\section{Conclusion}
Modelling free-surface flow problems using shallow-water equations are yet another field of research where physics-informed neural networks prove to be a suitable approach within the world of scientific machine learning.
We have demonstrated that choosing different representations of the underlying problem is crucial for the model's performance, and that going with an approach which might have some nice conservation property does not necessarily lead to better results empirically.
Furthermore, the used optimization methods to improve PINNs should be evaluated with care and might require hand-tuning to be useful for the problem at hand.

\section*{Acknowledgment}

Funded by Deutsche Forschungsgemeinschaft (DFG, German Research Foundation) under Germany's Excellence Strategy - EXC 2075 – 390740016. We acknowledge the support by the Stuttgart Center for Simulation Science (SimTech).

\bibliographystyle{IEEEtran}
\bibliography{IEEEabrv,iclr2021_conference}

\end{document}